# SD-AODV: A Protocol for Secure and Dynamic Data Dissemination in Mobile Ad Hoc Network

**Rajender Nath[1] and Pankaj Kumar Sehgal[2]**

**[1]Associate Professor, Department of Computer Science and Applications, Kurukshetra University, Kurukshetra, Haryana, India**

**[2]Assistant Professor, Department of Information Technology, MM Engineering College, MM University, Ambala, Haryana, India**

**Abstract**

Security remains as a major concern in the mobile ad hoc networks. This paper presents a new protocol SD-AODV, which is an extension of the exiting protocol AODV. The proposed protocol is made secure and dynamic against three main types of routing attacks- wormhole attack, byzantine attack and blackhole attack. SD-AODV protocol was evaluated through simulation experiments done on Glomosim and performance of the network was measured in terms of packet delivery fraction, average end-to-end delay, global throughput and route errors of a mobile ad hoc network where a defined percentage of nodes behave maliciously. Experimentally it was found that the performance of the network did not degrade in the presence of the above said attacks indicating that the proposed protocol was secure against these attacks.

## 1. Introduction

A multi-hop mobile ad hoc network (MANET) consists of a group of mobile wireless nodes that self configure to operate without infrastructure support. Network peers communicate beyond their individual transmission ranges by routing packets through intermediate nodes.

Security remains as a concern in MANET. In general, a MANET is vulnerable due to its fundamental cooperation of open medium, absence of central authorities, dynamic topology, distributed cooperation and constrained capability [1]. A node in the MANET without any adequate protection can become an easy target for attacks. Attacker just needs to be within radio range of a node in order to intercept the network traffic.

The attacks on MANET are classified as passive attacks and active attacks [22]. In passive attacks, an intruder snoops the data exchanged between the nodes without altering it. In these type of attacks, a selfish node abuses constrained resources such as battery power for its own benefit. The goal of an attacker is to obtain the information that is being transmitted that leads to the violation of massage confidentiality. Passive attacks are difficult to detect because the activity of the network is not disrupted in these attacks.

In active attacks, an attacker actively participates in disrupting the normal operation of the network services. These can be performed by injecting incorrect routing information to poison the routing table or by creating a loop. These attacks are further divided into external and internal attacks. External attacks are carried by nodes that are not authorized part of the network. Internal attacks come from compromised nodes, which are legitimate part of the network. Active attacks are very difficult to detect because the attacker is part of the network.

There are basically two approaches to securing a MANET: proactive and reactive. The proactive approach attempts to prevent security attack, typically through various cryptographic techniques. On the other hand, the reactive approach finds an attack and reacts accordingly. Both approaches has there own merits and suitable for different issues of security in MANET. Most of the secure routing protocols adopt proactive approach to securing routing control messages and reactive approach to secure data packet forwarding messages. A complete security solution requires both proactive and reactive approaches.

While a number of routing protocols [3-11] have been proposed by the Internet Engineering Task Force's MANET working group but they are silent in terms of security. Most of the MANET secure routing protocols have been proposed in the literature such as SEAD [12], ARIADNE [13], SAR [14], SRP [15], CONFIDANT [16], ENDAIRA [17], TESLA [21] etc. do not mitigate against these attacks. Some solutions against particular attacks have been presented by the researchers such as rushing attack and defenses [18], wormhole attack and defenses [19], sybil attack and defenses [20]. Because these solutions are designed explicitly with certain attack models in mind so they work well in the presence of designated attacks but may collapse under unanticipated attacks. Therefore, a more ambitious goal for MANET security is to develop a multifence security solution that can offer multiple lines of defenses against both known and unknown security threats.

Rest of the paper is structured as follows: Section 2 discusses the base routing protocol AODV. Section 3 describes the new protocol SD-AODV. Section 4 describes and compares the simulation experiment and result performed on AODV and SD-AODV protocol in presence of malicious nodes. Section 5 gives the concluding remarks.

## 2. Ad hoc On-demand Distance Vector (AODV) Routing Protocol

AODV is an improvement on DSDV [23] because it typically minimizes the number of required broadcasts by creating routes on a demand basis. AODV routing protocol uses reactive approach for finding routes, that is, a route is established only when it is required by any source node to transmit data packets. The protocol uses destination sequence numbers to identify the recent path. In this protocol, source node and the intermediate nodes store the next node information corresponding to each data packet transmission. In an on-demand routing protocol, the source node floods the Route REQuest (RREQ) packet in the network when a route is not available for the desired destination. It may obtain multiple routes to different destinations from a single RREQ. A node updates its path information only if the destination sequence number of the current packet received is greater than the last destination sequence number stored at the node.

A RREQ carries the source identifier (SrcID), the destination identifier(DestID), the source sequence number (SrcSeqNum) and destination sequence number (DestSeqNum), the broadcast identifier (BcastID), and the time to live (TTL) field. DestSeqNum shows the freshness of the route that is selected by the source node. When an intermediate node receives a RREQ, it either forwards it or prepares a route reply (RREP) if it has a valid route to the destination. The validity of a route at the intermediate node is determined by comparing the sequence number at packet. If a RREQ is received multiple times, which is indicated by BcastID-SrcID pair, then the duplicate copies are discarded. All intermediate nodes having valid routes to the destination, or the destination node itself are allowed to send RREP packets to the source. Every intermediate node, while forwarding a RREQ, enters the previous node address and its BcastID. A timer is used to delete this entry in case a RREP is not received before the timer expires. This helps in storing an active path at the intermediate node as AODV does not employ source routing of the data packets. When a node receives a RREP packet, information about the previous node from which the packet was received is also stored in order to forward the data packet to this next node as the next hop towards the destination.

## 3. Proposed SD-AODV Protocol

The existing AODV protocol is not secure against any routing attack. We have extended the existing AODV protocol to make it secure against the three types of routing attacks- Wormhole attack, Byzantine attack and Blackhole attack. The proposed protocol is named SD-AODV (Secure and Dynamic Ad Hoc On-Demand Distance Vector) which is secure and dynamic against In following paragraphs we describe different schemes to make the protocol secure against the above said three attacks.

Let N= $\{n_1, n_2, n_3, \ldots\ldots, n_k\}$ is a set of k nodes in the network that includes destination nodes and malicious node. Let D=$\{d_1,d_2,d_3,\ldots\ldots,d_j\}$ is a set j destination nodes where $D \subset N$ , j<k and M=$\{m_1, m_2, m_3,\ldots\ldots\ldots,m_h\}$is set of h malicious nodes where $M \subset N$ , h<k. Any member of M can act as a malicious node to perform either wormhole attack malicious node or byzantine malicious node or blackhole malicious node. The three different schemes has been formed and tested to safeguard against three different attacks.

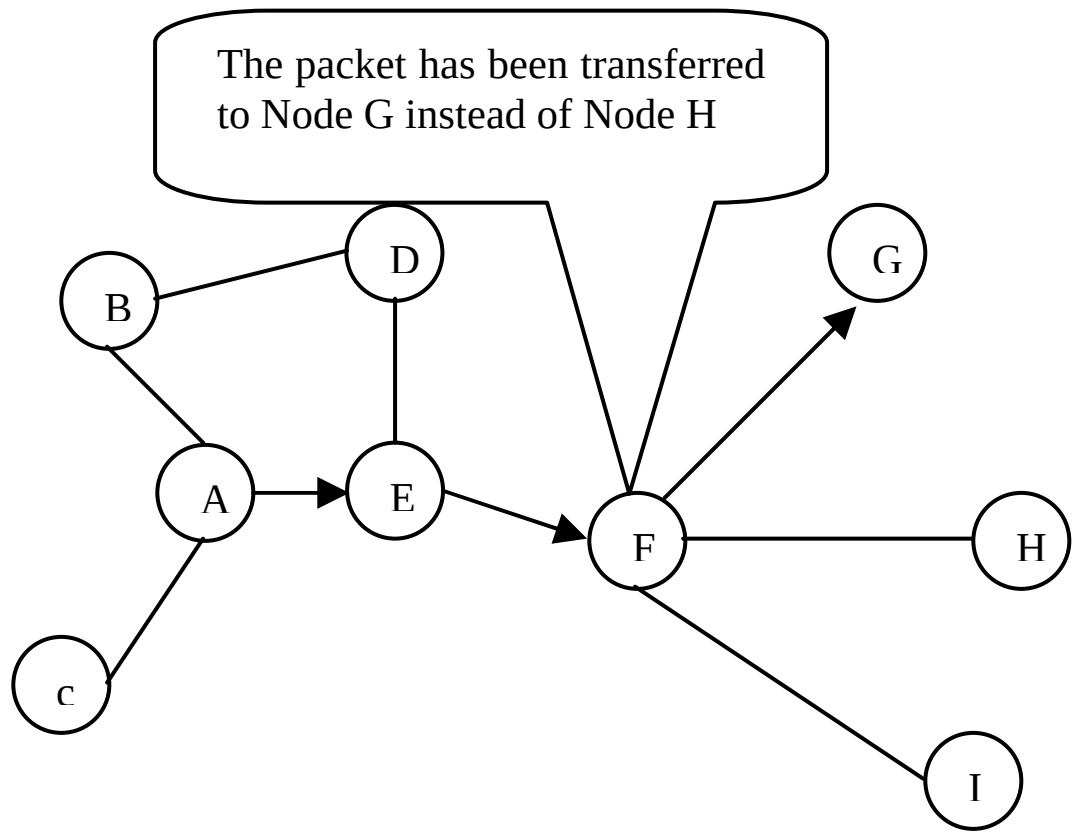

Figure 1. Wormhole Attack

The first scheme makes the proposed protocol secure against the wormhole attack. In the wormhole attack, an attacker receives packets at one location in the network and tunnels them to another location in the network. The tunnel between these two locations is referred as a wormhole. Due to the broadcast nature of the radio channel, the attacker can create a wormhole even for packets not addressed to itself. In Figure 1, let us assume node A is source node, H is destination node and E is malicious node, which can commit wormhole attack. When node A broadcasts a RREQ packet to its neighbors B, C and E, then the malicious node E commit wormhole attack and changes its destination address to G hence the request does not reach to the destination node H.

To safeguard against the wormhole attack we make use of hash chains in the proposed protocol. Hash chains are created by applying a secure hash algorithm [24]. This scheme is used for protecting the portion of the information in the RREQ messages, which is destination address. The protocol computes the digest (DigestAddr) of the destination address and appends that with the RREQ packet. It computes and compares the destination address at every intermediate node. If the destination address has been changed by the intermediate node, then it declares the node as malicious node and change the destination address to its original. Steps of the scheme are summarized below:

Step 1: Compute DigestAddr of destination address by calling
- SHAReset (…) function and
- SHAInput (…) function

Step 2: Append computed digest in DigestAddr of Data packet.

Step 3: if ($n_x$ is an Intermediate Node) then
- Fetch Destination address from Data packet and compute digest named New_computed_Digest by calling
- Call SHAReset (…) function and
- Call SHAInput (…) function

Step 4: Compare DigestAddr with new computed digest in Step 3
- if(DigestAddr=New_Computed_Digest) then
  - $m_z$ node is detected as malicious node
  - SD-AODV change destination address field to accurate destination address

The second scheme makes the proposed protocol secure against the byzantine attack. In the byzantine attack, a malicious node or a set of malicious node works in collusion and carries out attacks such as creating routing loops and routing packets on non-optimal paths. It consumes energy and bandwidth of the network. In Figure 2, let us again assume node B is source node, H is destination node and C is malicious node, which can commit routing loop attack. When node B wants to transmit a data packet to the destination node H, the malicious node C loops the information back to node A as shown in Figure 2 and the packet does not reach to the destination node H.

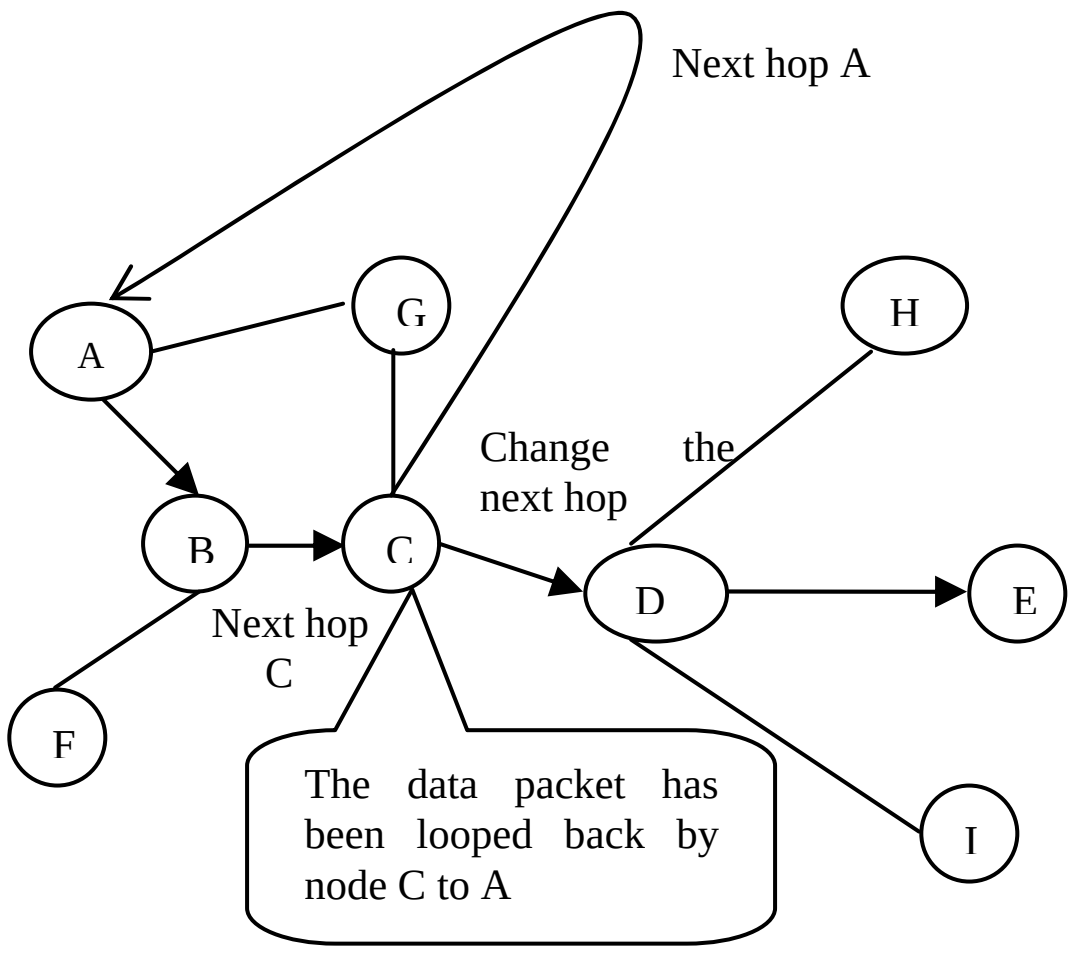

Figure 2. Byzantine Attack

SD-AODV protects the mutable information such as Next-hop field in the RREQ. This assures that nodes receiving AODV messages that the Next-hop value provided are accurate and have not been changed by the malicious node on the path. Every time when a node route packet to MAC layer the protocol dynamically save the value of Next-hop field. Whenever any intermediate node receives an RREQ message, the protocol verifies the Next-hop values. If the Next-hop value has been changed, the node is declared as a malicious node. The protocol then changes the Next-hop value to its original. Steps of the scheme are summarized below:

Step 1: Everytime when RoutePacketAndSendToMac(…) called
- SD-AODV save the value of Next-hop dynamically.

Step 2: if ($n_x$ is an Intermediate Node and $d_y$ exist in route table) then

If (Address of Next-hop has been altered) then
- $m_z$ node is detected as malicious node
- SD-AODV changes its value to original by updating route table.

The third scheme makes the proposed protocol secure against the blackhole attack. In the blackhole attack, a malicious node falsely advertises good paths (e.g. shortest path or most established path) to destination node during the route request phase. The intention of malicious node could be to hinder the route request phase or to stop all data packets being sent to the actual destination node. In Figure 3, let us again assume node A is source node, D is destination node and G is malicious node, which can commit routing blackhole attack. When node A wants to transmit a route request packet to the destination node H, the malicious node E advertise itself as shortest path and sends route reply shown in Figure 3 and the packet does not reach to the destination node H.

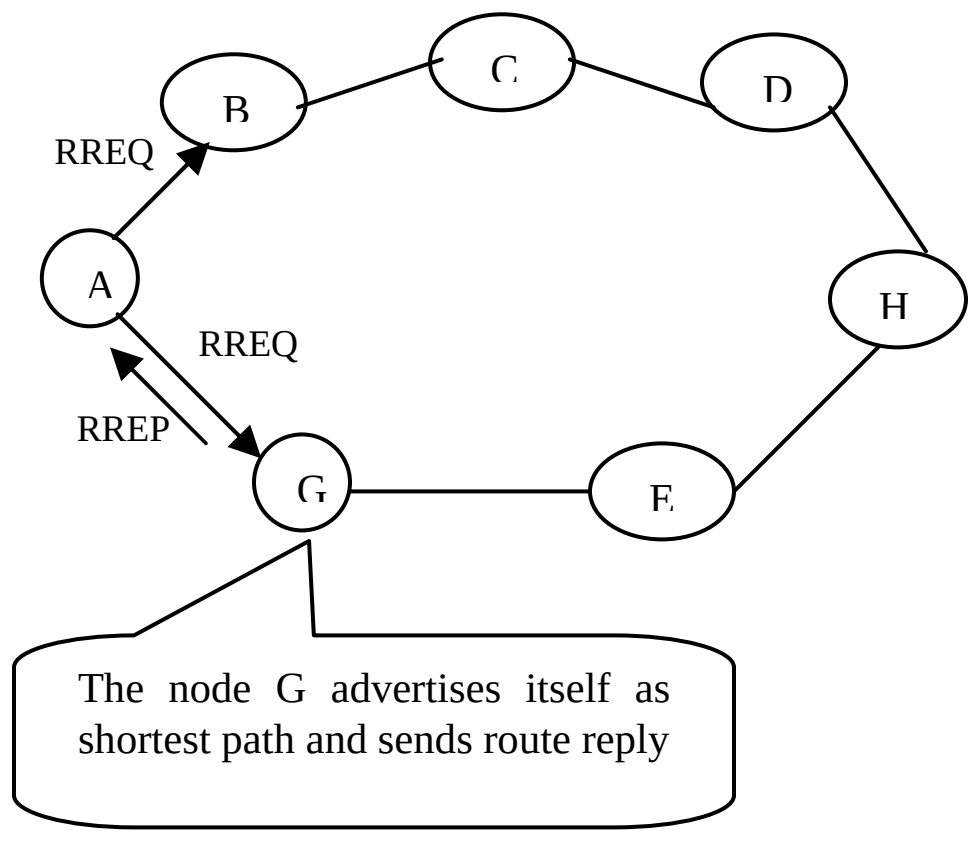

Figure3. Blackhole Attack

SD-AODV dynamically protects the route establishment process, by cross verifying the RREP (Route reply) message to ensure that the sender will not receive any RREP from malicious node. Steps of the scheme are summarized below:

Step 1: if ($n_x$ is a Intermediate Node) then
- Call RoutingAodvCheckRouteExist(…)
- if ($d_y$ does not exists in Route Cache) then
- $m_z$ node is detected as malicious node

Step 2: SD-AODV stop the route establishment process by calling
- RoutingAodvInitiateRREPbyIN (…)
- RoutingAodvRelayRREQ(…..)

## 4. Simulation Experiment and Results

A simulation experiment was performed by using Glomosim [2] simulator to study the effects of all the three attacks mentioned above on the proposed protocol: SD-AODV. The simulation experiment was performed on a computer with Intel core 2 Duo 1.7 GHz processor and 2GB RAM. The simulation experiment was performed twice by taking 50 and 100 node to study the effects of the three attacks by measuring the performance of the network. In each of the case i.e. 50 nodes and 100 nodes simulation was carried out several times with different seed values. Other parameters that were taken for simulation are shown in the table 1.

Table 1: Simulation Parameter Value

| Parameter | Vale | Description |
|---|---|---|
| Terrain Range | 1KM× 1KM | X,Y Dimension in Kilometer |
| Power Range | 250m | Nodes's power range in meters |
| Simulation Time | 100 s | Simulation duration in seconds |
| Node Placement | Uniform | Node placement policy |
| Mobility | 5m/s-20 m/s | Random Waypoint in meter per second |
| Traffic Modal | CBR | Constant bit rate |
| Packet Size | 512 Bytes | Minimum transfer unit |
| MAC | IEEE 802.11 | Medium Access Control Protocol |
| Bandwidth | 2MBPS | Node's Bandwidth in Mega bits per second |
| Routing Protocol | AODV | Base routing protocol for ad hoc networks |

The same experiment was repeated for the existing protocol AODV in order the compare it with the proposed protocol SD-AODV.

### 4.1 Testing SD-AODV and AODV Protocols against Wormhole, Byzantine and Blackhole attacks

The performance of the network was evaluated by using the following four metrics (i) Packet Delivery Fraction (PDF), (ii) Average End-to-End Delay, (iii) Throughput and (iv) Route error.

Packet Delivery Fraction (PDF): It is a ratio of the data packets delivered to the destinations to those generated by the Constant Bit Rate (CBR) sources.

Average End-to-End Delay: This includes all possible delays caused by buffering during route discovery latency, queuing at the interface queue, retransmission delays at the MAC, and propagation and transfer times.

Throughput: It is equal to the average performance of all nodes during simulation. It is a calculation of bits per second processed by each node.

Route Errors:The error messages garneted by the protocol during simulation.

Simulation was performed by taking different seed values. In the experiment, the numbers of malicious node were increased starting from 5% to maximum of 30 % in the step of 5%. The Glomosim simulator generated a GLOMO.STAT file which contained all the statistics regarding number of packets send, number of packets received, number of bytes sents, number of bytes received, throughput(bits per second), delay (in seconds), number of route errors etc.

### 4.1.1 Using Packet Delivery Fraction (PDF)

PDF was calculated by extracting data from the GLOMO.STAT file and four curves (one for wormhole attack on AODV, one for byzantine attack on AODV and one for blackhole attack on AODV and one for three attack on SD-AODV) are plotted by taking %age of malicious node on X-axis and %age of PDF on Y-axis as shown in Figure 4(a) and 4(b) for 50 nodes and 100 nodes respectively. From the Figure 4(a) and (b), it is quite clear that SD-AODV not only prevents from various attacks but also gives better performance while increasing the number of malicious nodes.

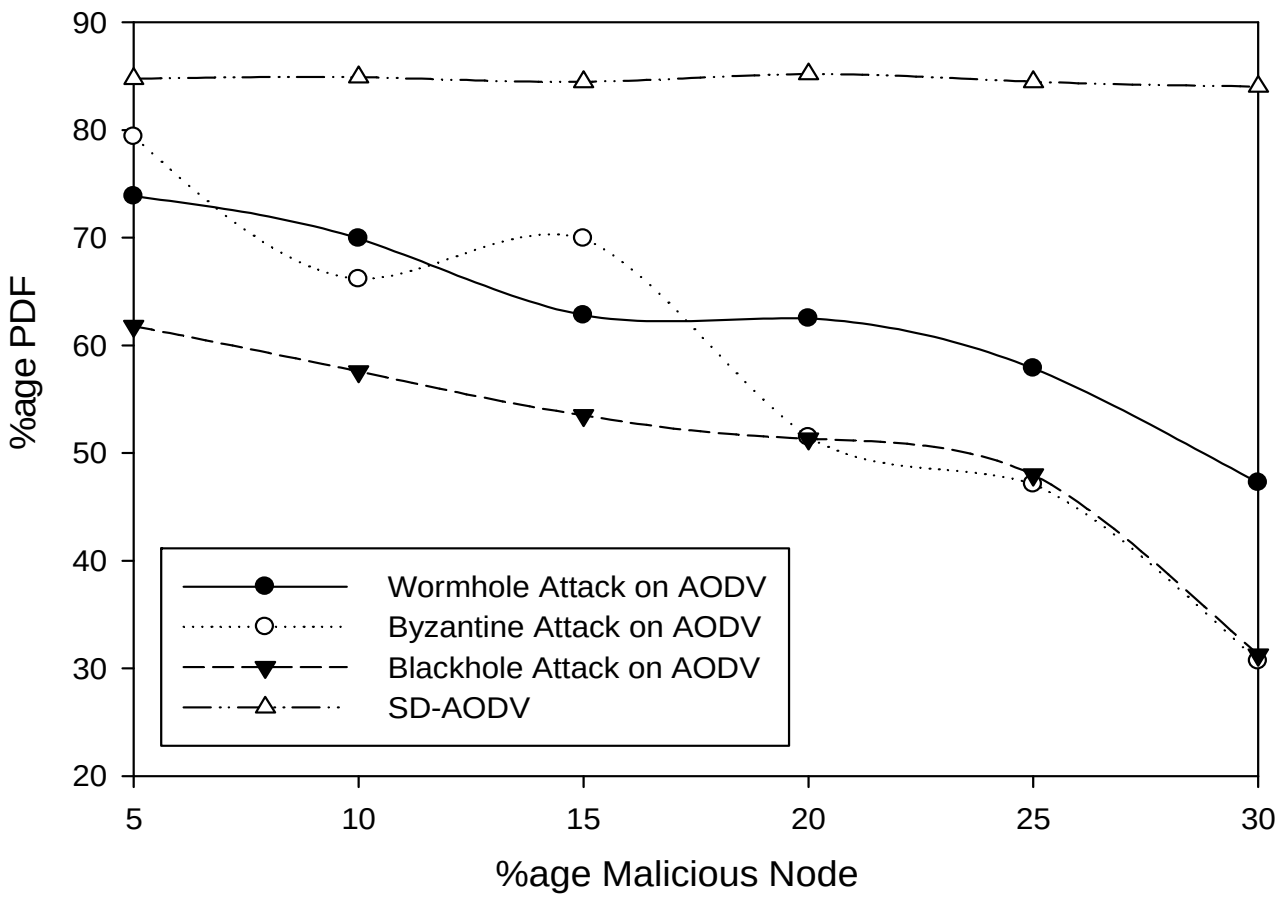

Figure 4(a). PDF Comparison of different attacks on AODV and SD-AODV in 50 nodes scenario

### 4.1.2 Average End-to-End Delay

Average end-to-end delay was calculated by extracting data from the GLOMO.STAT file and four curves (one for wormhole attack on AODV, one for byzantine attack on AODV and one for blackhole attack on AODV and one for three attack on SD-AODV) are plotted by taking %age of malicious node on X-axis and Average delay on Y-axis as shown in Figure 5(a) and 5( b) for 50 nodes and 100 nodes respectively. The average delay has been increased almost double in case of 100 nodes as compared to 50 nodes. From the Figure 5(a) and 5(b), it is quite clear that the average delay has increased in the case of SD-AODV due to overhead increased for protection against three attacks. But it is still less as compare to Byzantine attack on AODV because it loops back data and utilize resources and bandwidth

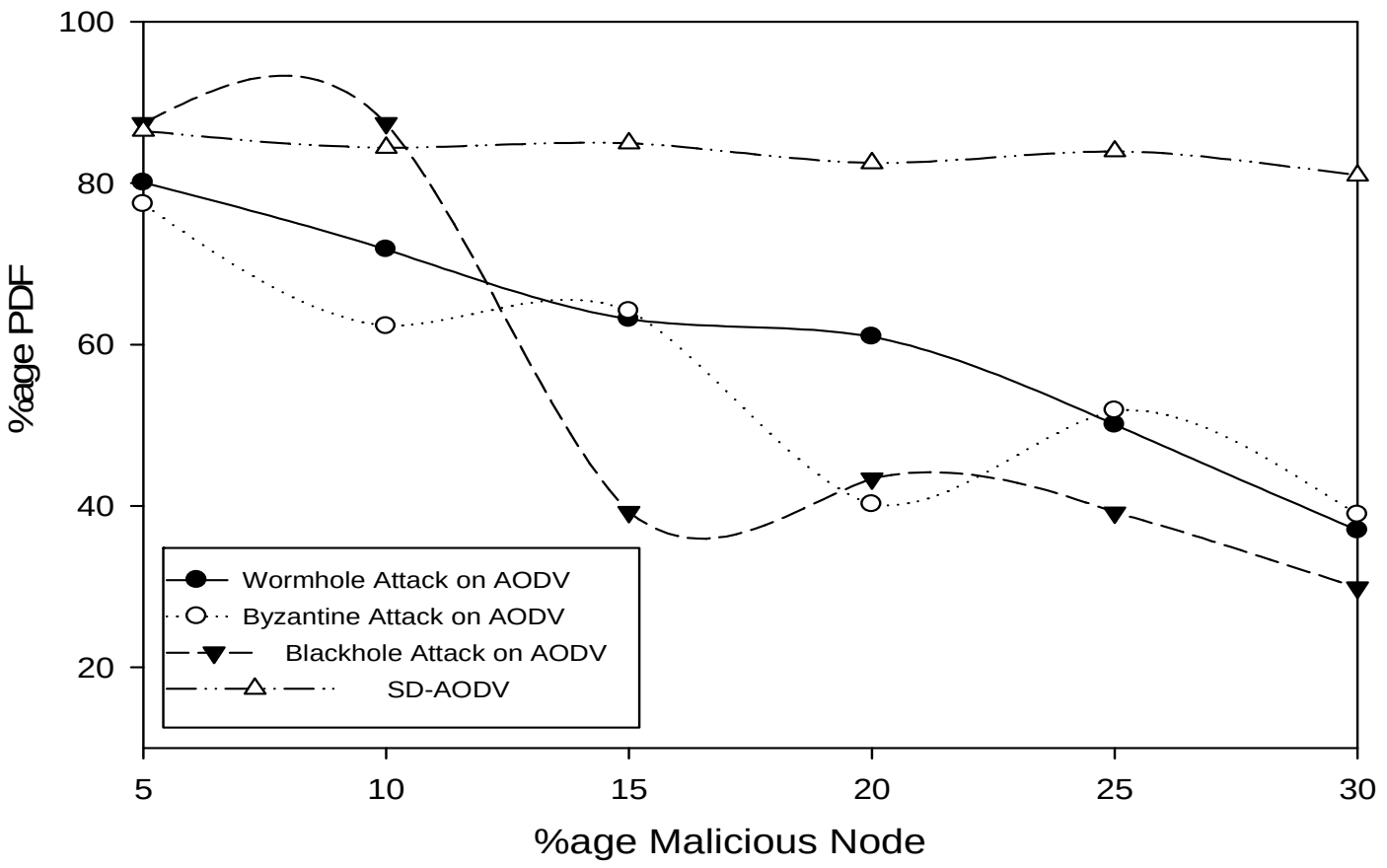

Figure 4(b). PDF Comparison of different attacks on AODV and SD-AODV in 100 nodes scenario

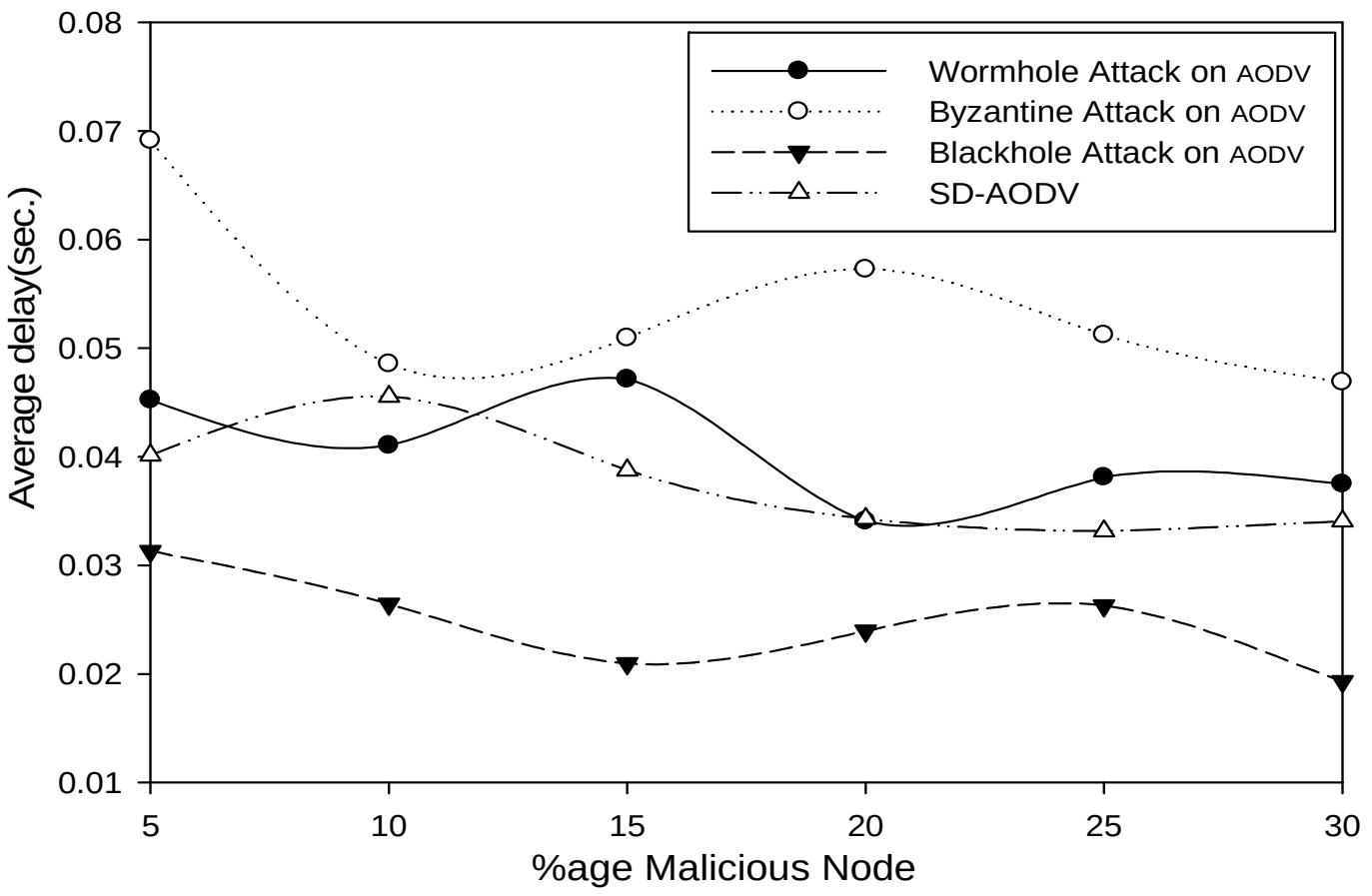

Figure 5(a). Average End-to-End Delay Comparison of different attacks on AODV and SD-AODV in 50 nodes scenario

### 4.1.3 Throughput

Throughput was calculated by extracting data from the GLOMO.STAT file and Four curves (one for wormhole attack on AODV, one for byzantine attack on AODV and one for blackhole attack on AODV and one for three attack on SD-AODV) are plotted by taking %age of malicious node on X-axis and throughput (bits per second) on Y-axis as shown in Figure 6(a) and 6(b) for 50 nodes and 100 nodes respectively. It is obvious from the figures that, throughput is increased in case of SD-AODV due to more calculation work for protective measures.

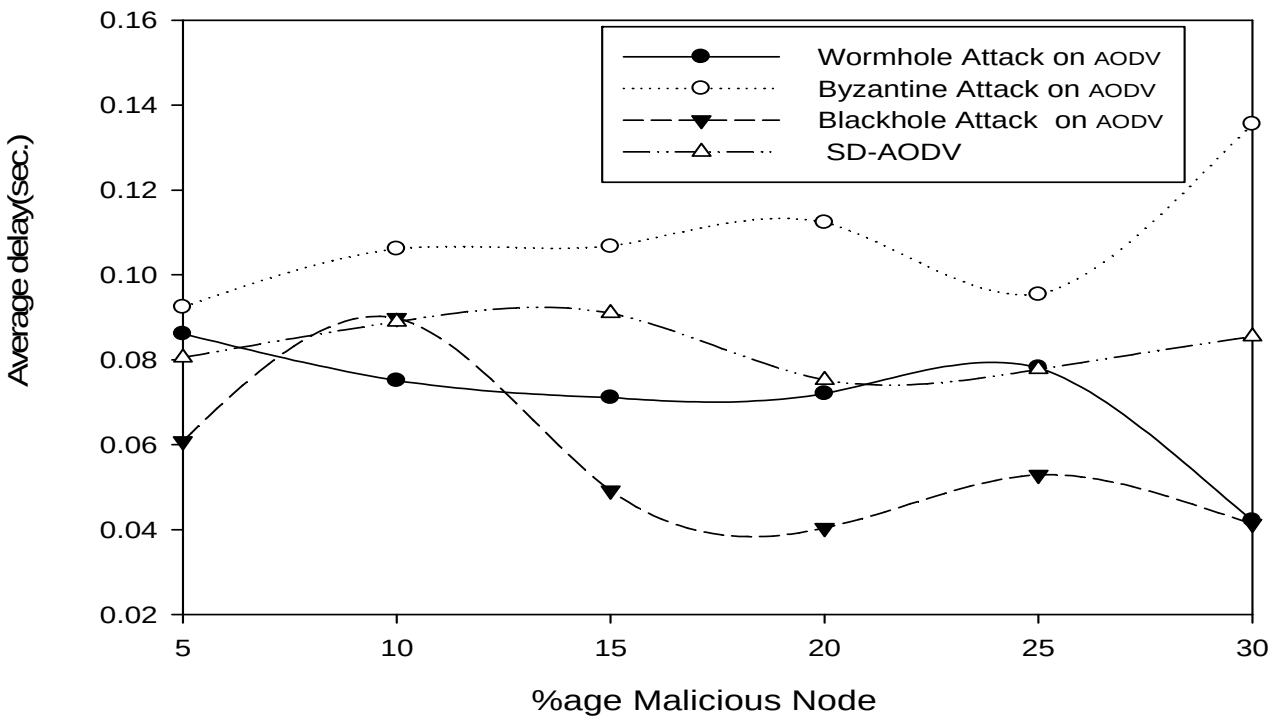

Figure 5(b). Average End-to-End Delay Comparison of different attacks on AODV and SD-AODV in 100 nodes scenario

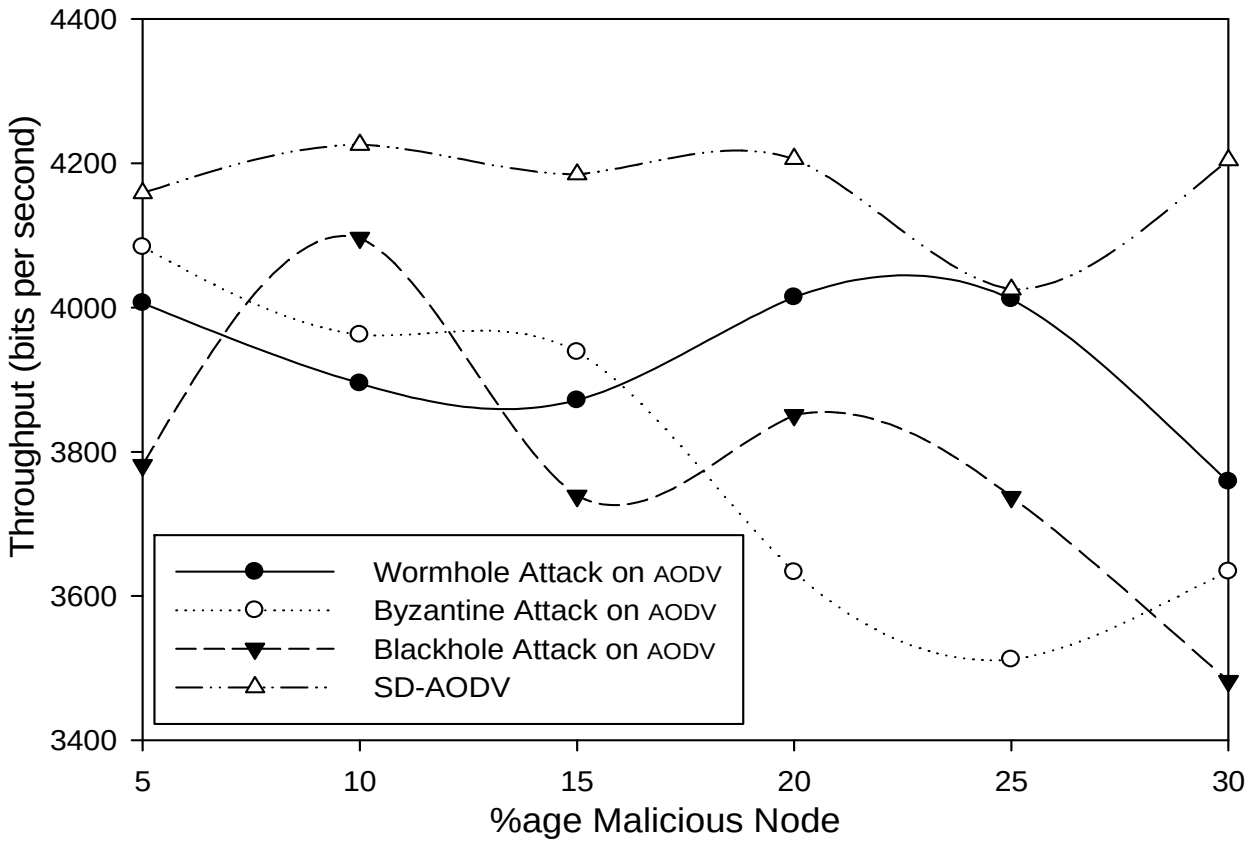

Figure 6(a). Throughput Comparison of different attacks on AODV and SD-AODV in 50 nodes scenario

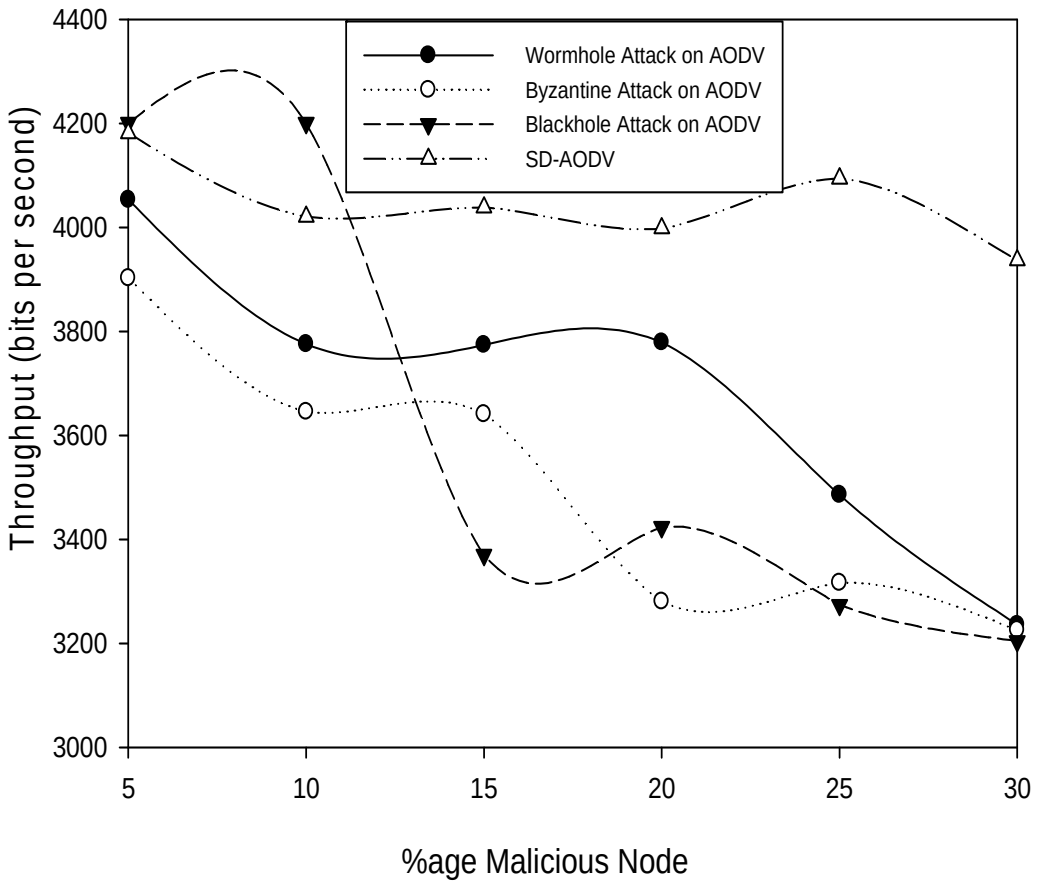

Figure 6(b). Throughput Comparison of different attacks on AODV and SD-AODV in 100 nodes scenario

## 4.1.4 Route Errors

Number of route errors was calculated by extracting data from the GLOMO.STAT file and Four curves (one for wormhole attack on AODV, one for byzantine attack on AODV and one for blackhole attack on AODV and one for three attack on SD-AODV) are plotted by taking %age of malicious node on X-axis and number of route errors on Y-axis as shown in Figure 6(a) and 6(b) for 50 nodes and 100 nodes respectively. It is quite clear that number of route errors have drastically increased in the case of blackhole attack because it attacks on route establishment process. Route errors are nominal in case of SD-AODV.

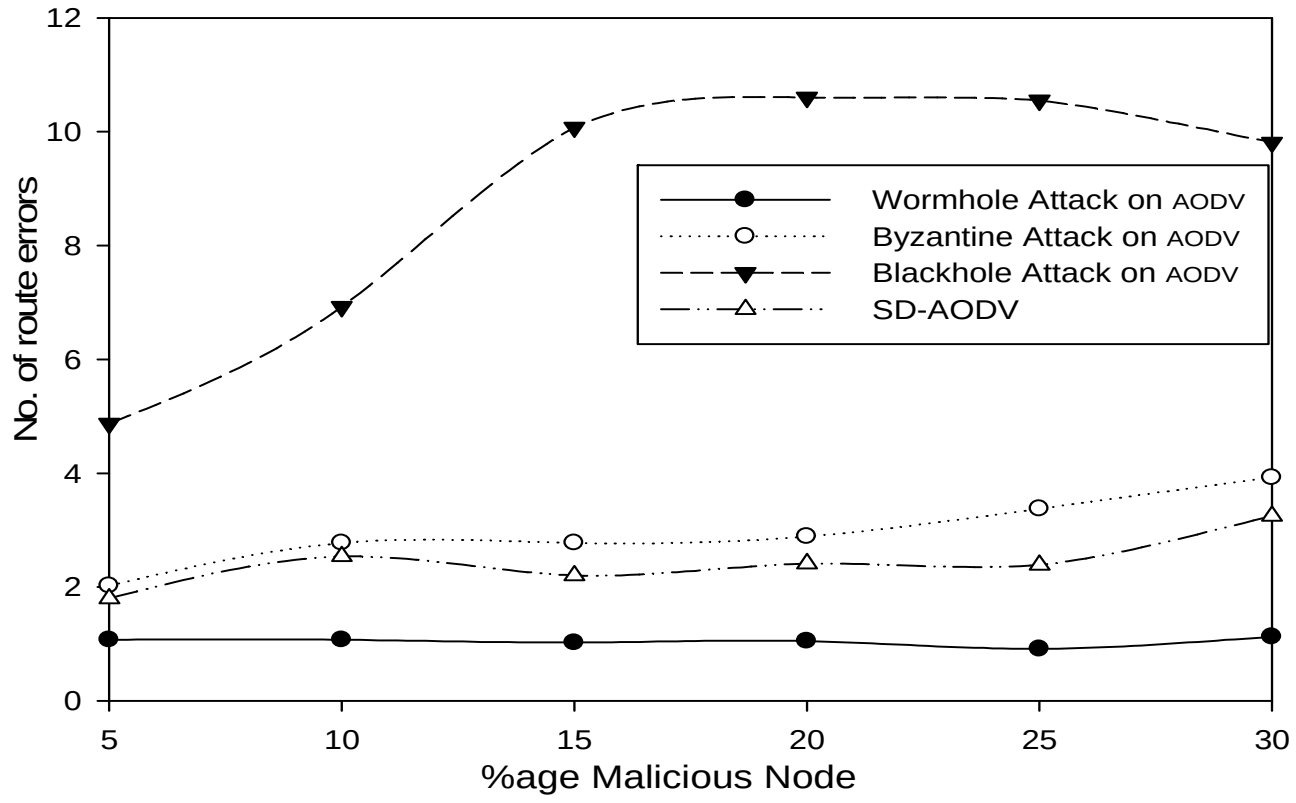

Figure 7(a). No. of Route Error Comparison of different attacks on AODV and SD-AODV in 50 nodes scenario

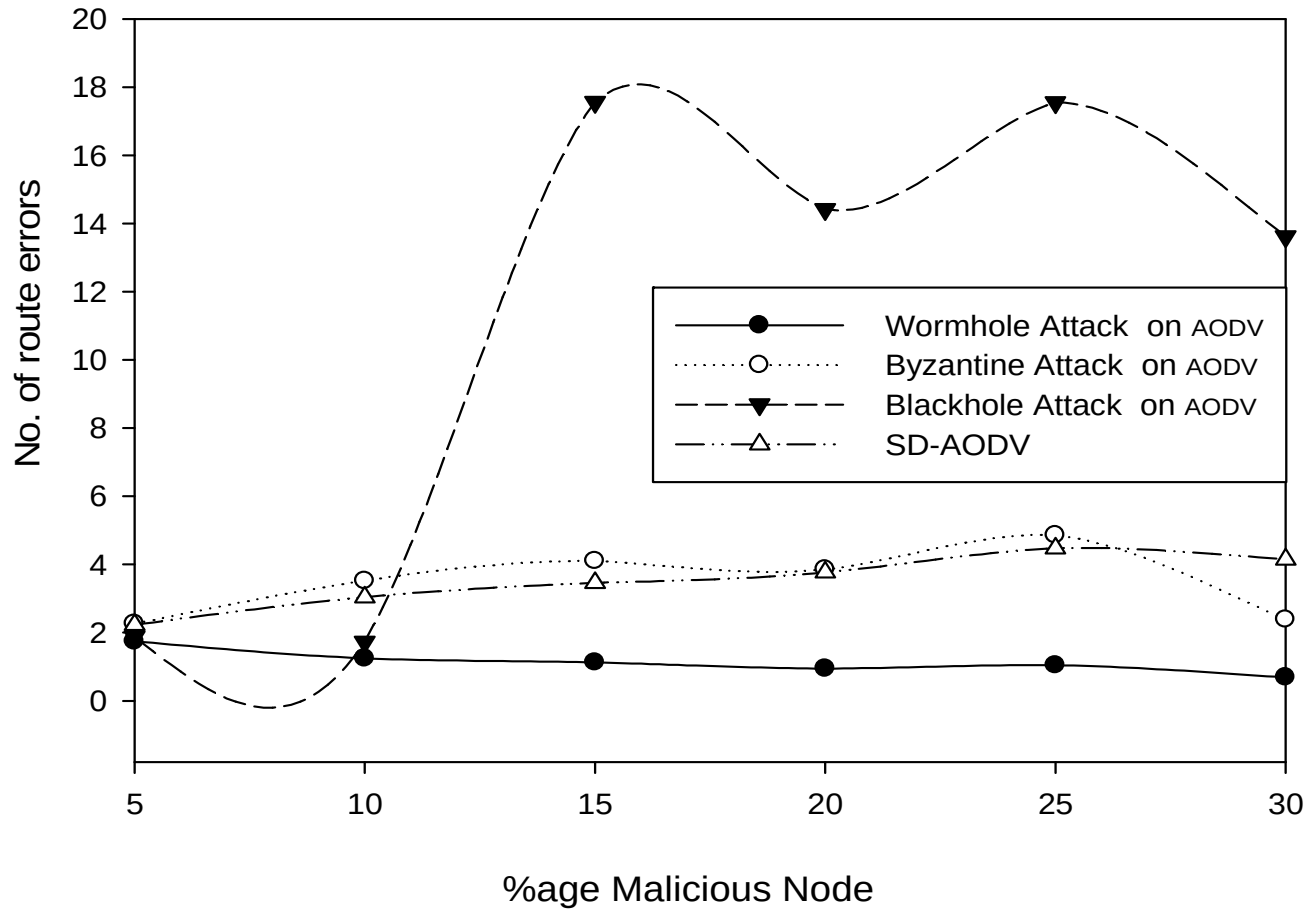

Figure 7(b). No. of Route Error Comparison of different attacks on AODV and SD-AODV in 100 nodes scenario

## 4. Conclusion

In this paper, we have presented a protocol which is secure & dynamic against three types of attacks - wormhole attack, Byzantine attack and blackhole attack. The protocol is based on the existing protocol AODV. The proposed protocol has been tested on Glomosim by using four metrics- packet delivery Fraction, average end-to-end delay of data packets, throughput and route error against widely used protocol AODV.

The packet delivery fraction (PDF) metric has shown that all the three routing attacks sharply decrease the PDF performance in the case of AODV protocol but in case of the proposed protocol SD-AODV there is no fall in the PDF, which has clearly indicated that the proposed protocol has became secure against the three attacks in question.

The Average end-to-end delay metric has shown that average delay has increased in the case of proposed protocol when the node has got the blackhole attack and wormhole attack, but it is lesser in case of Byzantine attack. The straight increase in delay can be attributed to the overhead incurred due the implementation of additional functionality of the SD-AODV protocol.

The result of throughput metric has shown that it is higher in case of the SD-AODV protocol as comparison to AODV protocol in the presence of the three attacks. This again indicates that the proposed protocol has become secure against these attacks.

The route errors have drastically decreased in case of SD-AODV protocol, but the results have also shown that the route errors are slightly more in case of SD-AODV as compared to AODV protocol when wormhole attack takes place because in wormhole attack the destination address is changed cleanly without affecting the route.

In nutshell, we can say that the proposed SD-AODV protocol has become secure and dynamic against these three attacks.